\documentclass[pra,twocolumn,showpacs]{revtex4-1}
\usepackage{graphicx}
\usepackage{amsmath}
\usepackage{color}
\usepackage{relsize}
\usepackage{hyperref}
\usepackage{gensymb}
\usepackage[utf8]{inputenc}

\begin{document}
	
	\title{Moir\'e super-lattice structures in kicked Bose--Einstein condensates}
	\author{L.~J.~O'Riordan, A.~C.~White and Th.~Busch}
	\affiliation{Quantum Systems Unit, Okinawa Institute of Science and Technology Graduate University, Onna, Okinawa 904-0495, Japan.}
	
	\begin{abstract}
	Vortex lattices in rapidly rotating Bose--Einstein condensates lead to a periodic modulation of the superfluid density with a triangular symmetry. Here we show that this symmetry can be combined with an external perturbation in order to create super-lattice structures with two or more periodicities. Considering a condensate which is kicked by an optical lattice potential, we find the appearance of transient moir\'e lattice structures, which can be identified using the kinetic energy spectrum.
	\end{abstract}
	
	\pacs{03.75.-b,67.85.-d}
	
	\maketitle
	
	
\section{Introduction}\label{sec:Introduction}
	    
Ultracold quantum gases have in recent years shown that they can be used as clean, cold and highly controllable simulators for fundamental effects in solid-state physics. Notable examples of these include strongly correlated lattice systems, Abrikosov vortex states, artificial gauge fields, and electronic spin-systems \cite{AO:Bloch_rmp_2008}. Here we add another example, by showing that super-lattice structures can be generated by simply applying an optical pulse to a vortex lattice state. This results in so-called moir\'e super-lattices which are known to appear in many solid-state \cite{nphys2272,SS:Romanova_jpcm_2011,SS:Li_scirep_2014}, and even biological~\cite{jns-moire}, systems.
		
Many of the effects observed in solid state physics owe their origin to the underlying periodicity of the system. The main effect of the periodicity is the appearance of a band-structure in the eigenspectrum, and designing and controlling this band-structure is of interest in order to control the physical behaviour. In ultracold systems periodicity can be most easily introduced by using optical lattice potentials~\cite{AO:Bloch_natphys_2012}. Made from a single optical wave-vector they provide geometries with perfect and simple periodicity, but combining lasers with two or more differing wave-vectors can allow for the creation of super-structures, such as double well lattices or more exotic geometries \cite{AO:Anderlini_jphysb_2006,AO:Barmettler_pra_2008,AO:Becker_njp_2010, AO:Jo_prl_2012}. Such super-lattices can, for example, be used to realise optical gauge fields  for neutral atoms~\cite{AO:Lin_nat_2009,AO:Gerbier_njp_2010,AO:Aidelsburger_prl_2011,AO:Georgescu_rmp_2014}. Given their experimental accessibility, optical lattices and super-lattices allow for studying interesting and often exotic new physics.	
More recently, spatially periodic systems (super-solids) have also been predicted to appear in the presence of extended interactions~\cite{SS:Boninsegni_rmp_2012}.

Using optical lattices on condensates which carry angular momentum is more complicated, as the rotation of the system is at odds with the (experimentally enforced) stationarity of the lasers that create the optical lattice. Assuming, however, that an optical lattice could be stationary in the co-rotating frame (see~\cite{AO:Tung_prl_2006} for an interesting example), several effects have been predicted in recent years. The pinning of vortices by a ramped optical lattice potential~\cite{AO:Reijnders_prl_2004,AO:Tung_prl_2006,AO:Mithun_prl_2014} shows that the vortices follow the lattice maxima, assuming a blue-detuned field. Increasing the optical lattice strength therefore causes the vortex lattice structure to undergo a phase transition from triangular to the geometry of the pinning lattice. 

In this work we propose a different approach to combining a rapidly rotating vortex lattice in a Bose--Einstein condensate with an optical lattice: switching the optical lattice on for a very short amount of time (such that only a phase is imprinted) creates a transient phononic excitation pattern of the same periodicity as the optical pulse. The combination of this excitation pattern, and that from the vortex lattice, generates a moir\'e super-lattice pattern in the condensate density \cite{BK:Amidror_2009}. Given the experimental difficulties in rotating the optical lattice with the condensate, the ability to imprint a phase pattern onto the condensate from a pulsed optical potential is much more realisable. 
		
The manuscript is structured as follows. In Section~\ref{sec:Model} we introduce the system of a Bose--Einstein condensate under rapid rotation and describe the kicking mechanism. The dynamics originating from this are discussed in Section~\ref{sec:kickvl} and the observed super-structures are explained using moir\'e interference theory. We conclude in Section~\ref{sec:Conclusions}.


\section{Model}
\label{sec:Model}
We consider a single component Bose--Einstein condensate in a harmonic trap, which in the mean-field limit can be described by the standard Gross--Pitaevskii Hamiltonian
\begin{equation}\label{eqn:gpe_h0}
	H_\text{GP} = -\frac{\hbar^2}{2m}\nabla^2 + \frac{1}{2}m\omega^2\mathbf{r}^2 + g\vert\Psi(\mathbf{r},t)\vert^2.
\end{equation}
Here $\omega$ is the trapping frequency, $g$ describes the strength of the two-particle interaction, and $m$ is the atomic mass. Imposing an external rotation around the $z$-axis and going into the co-rotating frame leads to a Gross--Pitaevskii equation of the form
\begin{equation}\label{eqn:gpe}
	i\hbar\frac{\partial}{\partial t}\Psi(\mathbf{r},t) = \left[ H_\text{GP}  -  \Omega_z L_z\right] \Psi(\mathbf{r},t),
\end{equation}
where $\Omega_z$ is the rotation frequency around the $z$-axis and $L_z = xp_y - yp_x$ is the angular momentum operator. For a rotational frequency close to the transverse trapping frequency, $\omega_{\perp}$, the condensate is known to respond by forming a triangular lattice of vortex lines in the $x$\nobreakdash-$y$~plane~\cite{Vtx:AboShaeer_sci_2001}. Since we are only interested in the effects stemming from the lattice ordering, we will choose the trapping frequency along the $z$-axis, $\omega_z$, to be much larger than the one along the transverse axis, $\omega_\perp$, so that the condensate assumes a pancake-shaped geometry~\cite{BEC:Fetter_revmodphys_2009}. This allows us to simplify our analysis by neglecting the $z$ dimension ($\textbf{r}\equiv [x,y]$), and treat the condensate as a two-dimensional system.
	
\begin{figure}[tb]
			\includegraphics[width=0.5\textwidth, angle=0, trim=0 0 0 0]{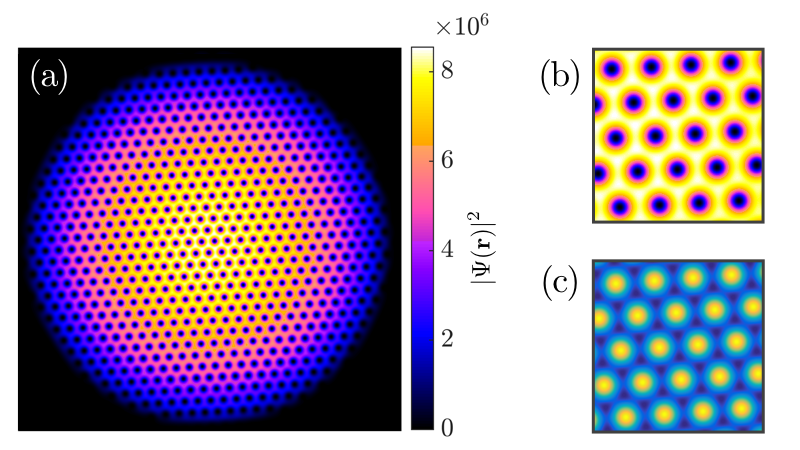}
			\caption{(a) Vortex lattice ground-state in a harmonic trap with $\omega_\perp=2\pi$ s$^{-1}$ and rotating at $\Omega=0.995\omega_\perp$. This plot shows a condensate with a diameter of approximately $580~\mu\textrm{m}$; (b) Zoom in of vortex lattice at central density; (c) Optical lattice potential with a periodicity matching that of the vortex lattice.}
			\label{fig:vlatt_gnd}
		\end{figure}

In the following we will solve eq.~\eqref{eqn:gpe_h0} numerically, using a GPU implemented Fourier split operator method \cite{NUMERICS}.
All simulations will use $N=4.9\times 10^5$ atoms of $^{87}$Rb, which have an  s-wave scattering length of $a_s=4.76\times10^{-9}$ m~\cite{BEC:Roberts_prl_1998}. The trapping frequency will be fixed at $\omega_{\perp}=2\pi$ s$^{-1}$, the rotation frequency at $\Omega=0.995\omega_\perp$, and the effective two-dimensional interaction strength is given by 
	\begin{equation}	
		g_\text{2d} = \frac{4\pi\hbar^2 a_s N}{m}\sqrt{\frac{m\omega_z}{2\pi\hbar}}.
	\end{equation}
The numerically calculated ground-state for the above set of parameters is shown in Fig.~\ref{fig:vlatt_gnd} and it can be characterised by the filling factor (or filling fraction) $\nu=N/N_v$, i.e.~the ratio of atoms to vortices~\cite{BEC:Fetter_revmodphys_2009}. For the chosen parameters, the number of vortices within the visible density region is approximately 600, giving a filling factor of $\nu \approx 800 $. This places the system within the mean-field quantum Hall regime, and therefore a description using Gross--Pitaevskii theory is adequate~\cite{Vtx:Schweikhard_prl_2004}. 
The vortex lattice is highly periodic and the triangular lattice vectors in $\mathbf{r}$-space are given by $\mathbf{a}_1 = a_v\{1,0\}$ and $\mathbf{a}_2 = a_v\{-1/2, \sqrt{3}/2\}$, where $a_v$ is the distance between vortex cores. For our analysis we chose to ignore vortices close to the condensate edge, as the distortions in the vortex lattice mean the system can no longer be modelled as a homogeneous arrangement of vortices. The reciprocal ($\mathbf{k}$-space) lattice vectors corresponding to the given $\mathbf{r}$-space vectors are $\mathbf{b}_1 = 4\pi/(\sqrt{3}a_v)\{\sqrt{3}/2,1/2\}$ and $\mathbf{b}_2 = (4\pi/(\sqrt{3}a_v))\{0,1\}$.
	
To create a perturbation the condensate is {\it kicked} by switching on an additional optical potential for a period of time that is much shorter than the rotation period of the Abrikosov lattice ($\tau_{\text{kick}}=10^{-5}$~s). The vortex lattice is therefore effectively stationary during the kick. In order to create effects based on periodicity, we chose the optical potential to have the form of a standing lattice, which has the same geometry as the Abrikosov lattice, however not necessarily the same orientation (see Fig.~\ref{fig:vlatt_gnd}(c)). To construct such a lattice potential, $V_{\text{opt}}$, we sum counter-propagating laser beams to get
	\begin{equation}
		V_\text{opt} = V_0\displaystyle\sum_{j}\cos^2 \left[ \textbf{k}_{j}\cdot\textbf{r} \right],
	\end{equation}
	where $V_0$ is the amplitude of the optical lattice potential, and $j=\lbrace 0,1,2,\ldots \rbrace$ is the index of each respective laser with a differing $\mathbf{k}$-space wave-vector. The triangular structure of the vortex lattice can then be matched by choosing wave-vectors corresponding to the optical potential that follow the vortex lattice vectors $\mathbf{b}_{1,2}$ and adding a third wave-vector with $\mathbf{k}_3 = 4\pi/(\sqrt{3}a_o)\{\sqrt{3}/2,-1/2\}$. Note that these wave-vectors have a lattice constant $a_o$, which is based on the optical intensity in order to compare with the vortex lattice later. 
 However, as the inter-vortex separation in atomic Bose--Einstein condensates is large, one needs to employ optical lattices with wavelengths on the order of tens of microns \cite{BEC:Fallani_optexp_2005,BEC:Williams_optexp_2008}.
For short kicks, and amplitudes on the order of $10^{-2} \mu $, where $\mu$ is the chemical potential of the system, the effect of the kick is limited to a phase imprint on the condensates wavefunction \cite{Vtx:Dobrek_pra_1999}, which subsequently leads to the development of a flow originating from the position of each maximum of the optical potential. In the following we will show that this in turn leads to well-defined phonon interferences, which, when overlaid on the periodically arranged vortex cores,  gives rise to the appearance of moir\'e super-lattice structures \cite{mor:murata_acsn_2010} in the condensate density. While in many solid state systems, for example graphene on hexagonal boron nitride~\cite{nphys2272}, the moir\'e structures are static super-structures, in our case they are dynamical and appear at well defined intervals.

	
To identify the moir\'e lattices, we perform a spectral decomposition of the kinetic energy of the condensate~\cite{CT:Nore_prl_1997,CT:Nore_pof_1997,CT:Bradley_prx_2012}. For this we write the wavefunction in terms of its density, $\rho(\mathbf{r},t)$, and 
phase, $S(\mathbf{r},t)$, as
$
		\Psi(\mathbf{r},t) = \sqrt{\rho(\mathbf{r},t)}\exp{\left[\mathrm{i}S(\mathbf{r},t)\right]},
$
and obtain from the kinetic energy term in the Gross--Pitaevskii energy functional
	\begin{equation}
		E_\text{kqp} = \int d\mathbf{r} \left( \frac{\hbar^2}{2m}| \nabla\sqrt{\rho(\mathbf{r},t)} |^2  + \frac{m}{2}|\sqrt{\rho(\mathbf{r},t)}\mathbf{v}(\mathbf{r},t) |^2\right).
	\end{equation}
The first term in this expression can be interpreted as the quantum pressure term, and
the second describes the kinetic energy. We denote the density weighted velocity field as $\mathbf{u}(\mathbf{r},t) = \sqrt{\rho(\mathbf{r},t)}\mathbf{v}(\mathbf{r},t)$, and further decompose it into compressible, $\mathbf{u}^c(\mathbf{r},t)$, and incompressible, $\mathbf{u}^i(\mathbf{r},t)$ terms
	\begin{equation}
		\mathbf{u(r},t) = \mathbf{u}^c(\mathbf{r},t) + \mathbf{u}^i(\mathbf{r},t),
	\end{equation}
defined by $\nabla \times \mathbf{u}^c(\mathbf{r},t) = 0$, and $\nabla \cdot \mathbf{u}^i(\mathbf{r},t) = 0$. This decomposition allows the energy contribution from vortex cores (incompressible) and phonon modes (compressible) to be separated~\cite{CT:Horng_pra_2009}. The compressible and incompressible kinetic energy spectra, $E^{c,i}(k)$, are calculated by averaging over shells in $\mathbf{k}$-space as~\cite{CT:Bradley_prx_2012}
	\begin{equation}
		E^{c,i}(k) = \frac{mk}{2}\sum\limits_{j\in\mathbf{r}} \int\limits_{0}^{2\pi}d\phi_k \frac{ |\mathcal{U}_j^{c,i}(\mathbf{k},t) |^2}{s_k},
	\end{equation}
	where
	\begin{equation}
		\mathcal{U}_j^{c,i}(\mathbf{k},t) = \int d^2 \mathbf{r} e^{-i(\mathbf{k}\cdot\mathbf{r})} u_j^{c,i}(\mathbf{r},t).
	\end{equation} 
	Here the $u_j^{c,i}(\mathbf{r},t)$ are the values of the position-space kinetic energy components in the specified shell, $\phi_k$ is the polar angle, and $s_k$ represents the number of elements in the chosen shell.
	
	
\section{Dynamics following a kick}\label{sec:kickvl}
	\subsection{Non-rotating condensate}

To understand the effect kicking has on a condensate carrying a vortex lattice, we will first briefly examine the situation where the vortex lattice is absent, i.e.~where there is no external rotation. In order to compare this situation with a rapidly rotating condensate, we adjust the trapping frequency in this section such that the background densities match.

\begin{figure}[tb]
			\includegraphics[width=0.48\textwidth]{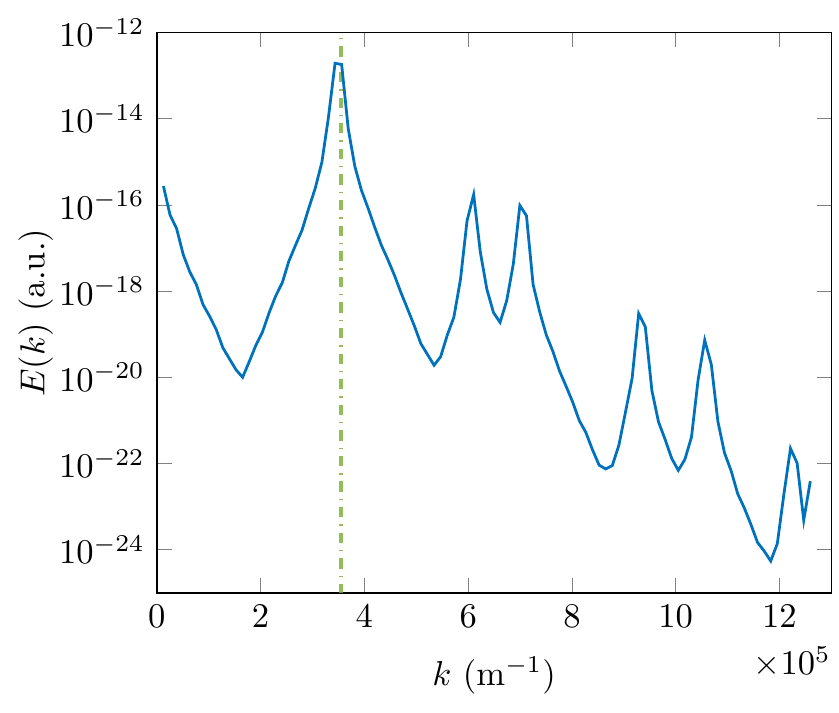}
			\caption{Compressible energy spectrum of a non-rotating condensate directly following a kick. A peak at $k=4\pi/(\sqrt{3}a_o)$ can be seen, which corresponds to the  lattice spacing, $a_o$ (indicated by the dashed line), and the smaller, higher energy peaks can be attributed to higher harmonics between nearest and next-nearest neighbours. }
			\label{fig:ekc_eki_novtx}
		\end{figure}

For a stationary (non-kicked) condensate the kinetic energy spectrum is constant during time-evolution and a kick leads to the appearance of new, time varying components. To observe this we numerically evolve the system by setting $V(\mathbf{r},t) = V_\text{ext}(\mathbf{r}) + V_\text{opt}(\mathbf{r},t)$, where the time dependent optical potential is only active for $\tau=10^{-5}$ s of the simulation time, and examine the compressible kinetic energy spectrum following the kick (see Fig.~\ref{fig:ekc_eki_novtx}). Unsurprisingly, the spectrum is dominated by a peak corresponding to the wave-vector associated with the optical potential at $k=4\pi/(\sqrt{3}a_o)$, and several smaller ones corresponding to its higher harmonics and higher harmonics of next nearest neighbour components of the lattice. As no rotation is created by the imparted phonon modes, the incompressible energy is negligible and we will therefore restrict our analysis to the compressible part of the spectrum.

\begin{figure}[tb]
			\includegraphics[width=0.48\textwidth]{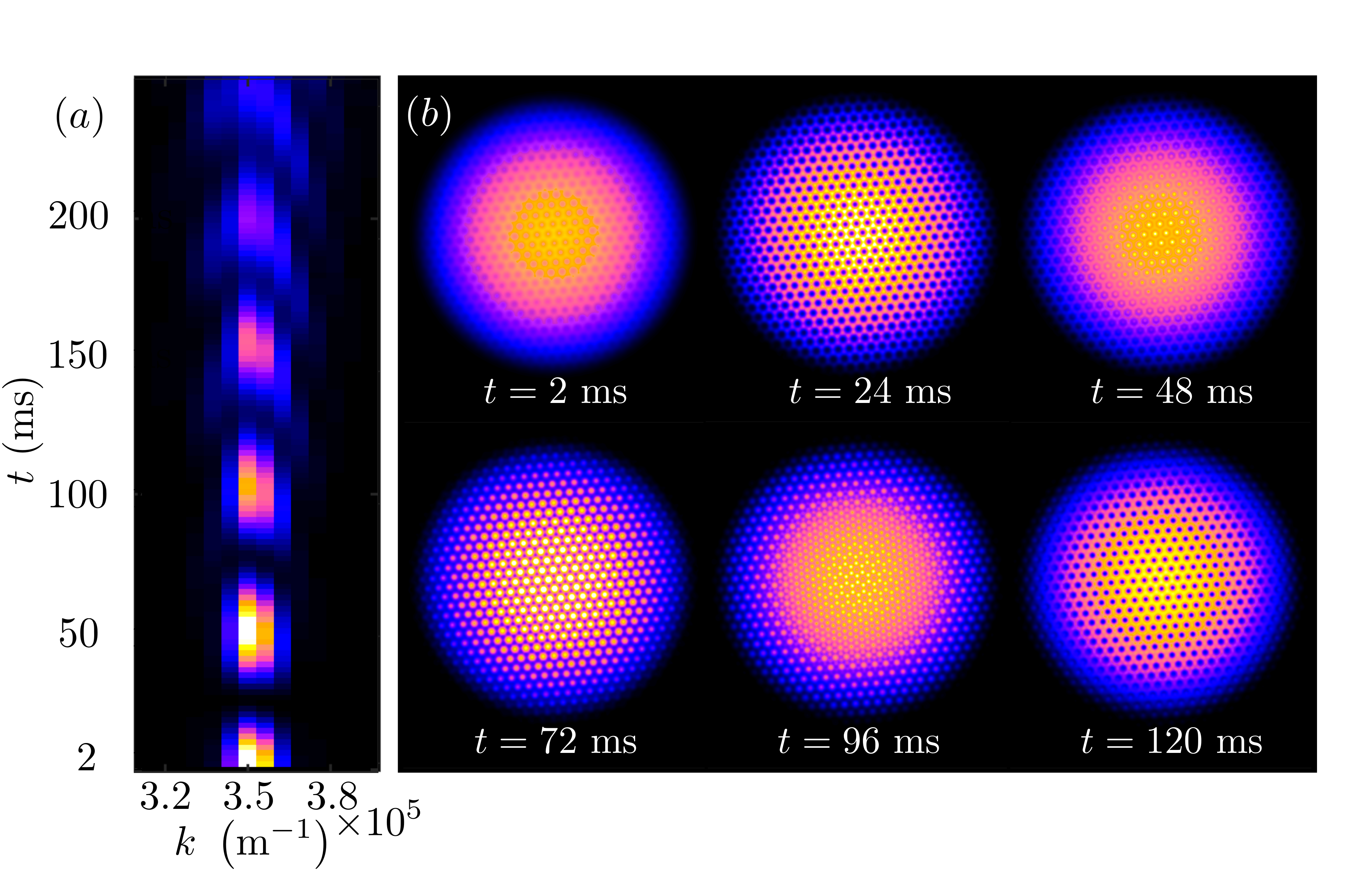}
			\caption{(a) Main peak of the compressible kinetic energy spectrum for a kicking strength of $V_0 \approx 1.35\times10^{-2}\mu$. It can be seen to revive, and eventually disperse over a wide range of wave-numbers.  (b) Condensate densities at several times during the evolution. A pattern matching the optical potential can be observed to appear and disappear several times over the course of the evolution.}
			\label{fig:novtx_p5k}
		\end{figure}

The evolution of the main peak in the compressible kinetic energy spectrum during the first 250 ms after the kick is shown in Fig.~\ref{fig:novtx_p5k}(a). It initially oscillates in and out of existence and eventually disperses over a wide range of wave-numbers. Snapshots of the density evolution are shown in Fig.~\ref{fig:novtx_p5k}(b), which clearly show that the oscillations correspond to the existence of a transient lattice pattern with several revivals, which has the same underlying structure as the optical potential. In fact, the lattice pattern is best formed whenever the main peak in the kinetic energy spectrum goes to zero, i.e.~when the imprinted kinetic energy has been converted into density modulations. The period of the oscillations can be related to the speed of sound divided by the lattice constant and therefore the appearance of the lattice can be attributed to phonon interferences. 

	\subsection{Rapidly rotating condensate}
	
Kicking a condensate carrying an Abrikosov vortex lattice with the above optical lattice gives an additional parameter, $\theta_\Delta$, which describes the orientation of the imprinted phonon lattice angle relative to the vortex lattice. For simplicity, we assume that the vortex and optical potential lattices have the same lattice constant, $a_v=a_o=a$ (see below for a discussion of the incommensurate case), which means that symmetry allows us to restrict the angle to $\theta_\Delta\in[0,\pi/3]$. We will show in the following that adjusting $\theta_\Delta$ leads to the appearance of different, transient super-structures in the condensate density.
			
\begin{figure}[tb]
			\includegraphics[width=0.48\textwidth]{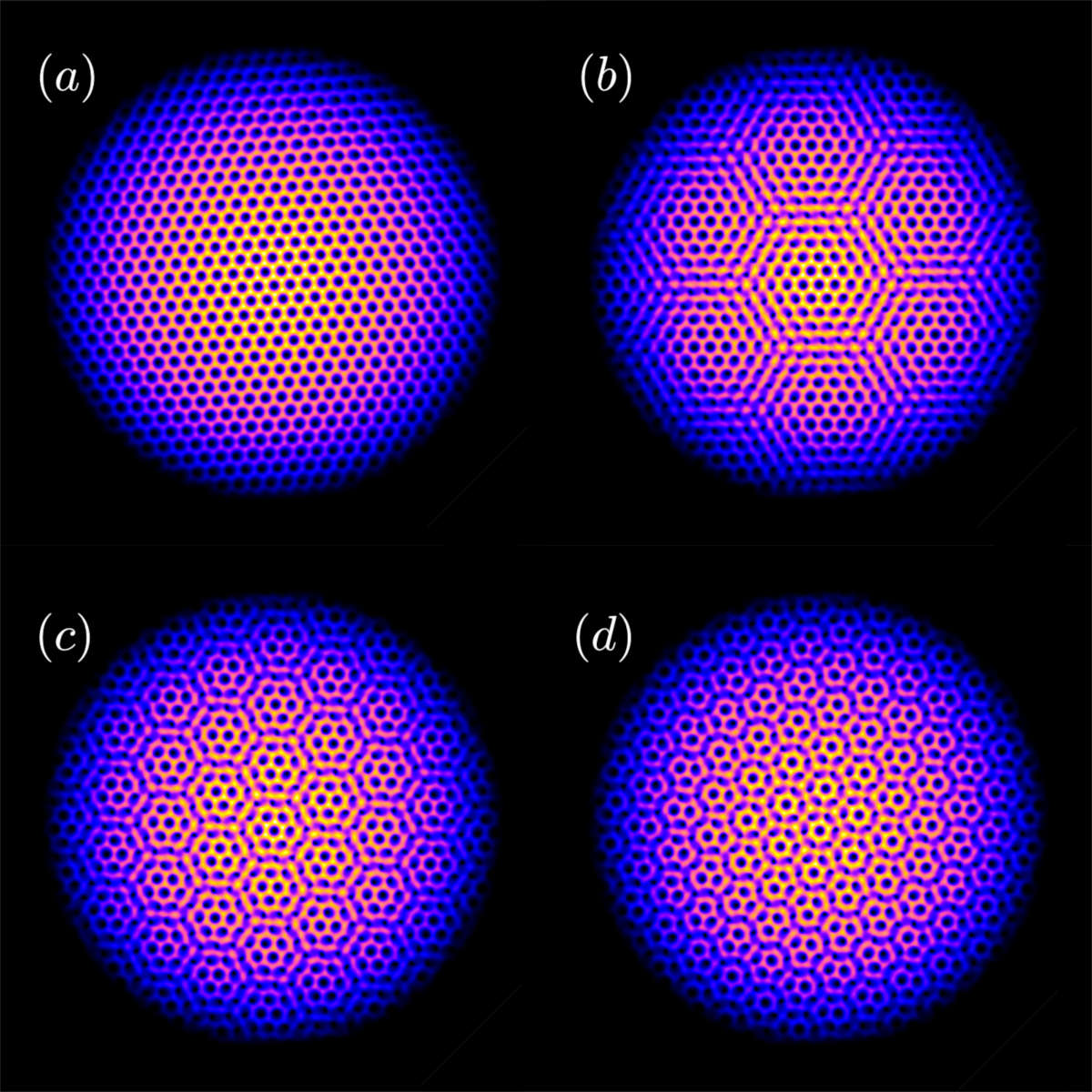}
			\caption{Condensate density at $t=1.4\times10^{-2}$ s for several optical lattice rotation angles. The cell size of the super-lattice structures can be seen to shrink as the angle is increased. The angles for the examples shown are $(a)~\theta_\Delta=0$, $(b)~\theta_\Delta=2\pi/45$, $(c)~\theta_\Delta=4\pi/45$, $(d)~\theta_\Delta=2\pi/15$. }
			\label{fig:moire_density}
		\end{figure}

If $\theta_\Delta=0$ (see Fig.~\ref{fig:moire_density}(a)) the kicking imparts kinetic energy at wave-numbers that are already well defined in the lattice. This leads to an expansion and contraction of the vortex cores in density only, and no significant change to the compressible kinetic energy spectrum. 

\begin{figure}[bt]
			\includegraphics[width=0.48\textwidth]{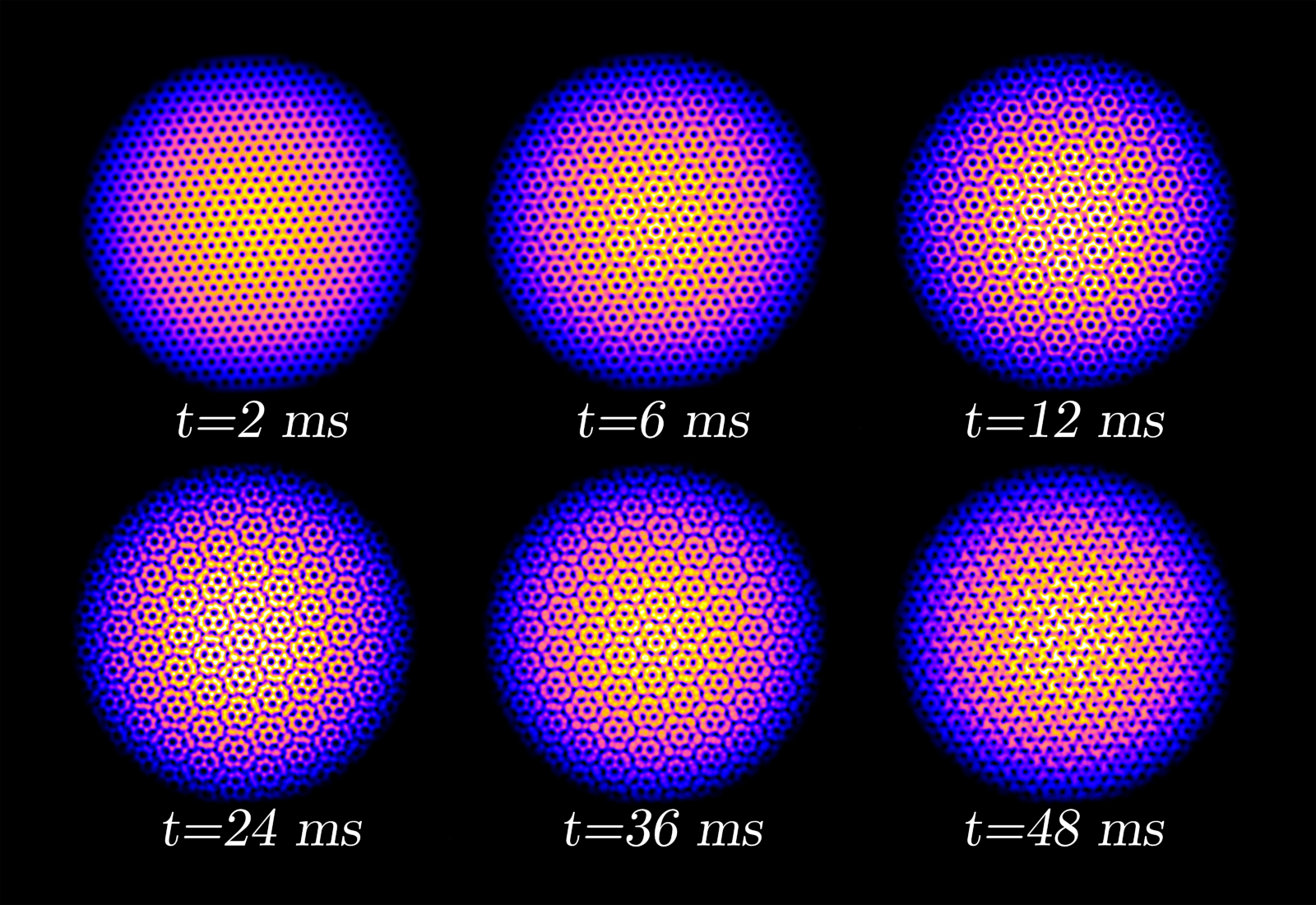}
			\caption{Condensate density after receiving a kick with $\theta_\Delta=\pi/9$. The appearance and disappearance of a moir\'e structure with wavelength $\lambda_M \approx 2.9 a$ over a timescale of about $50$ ms can be seen.}
			\label{fig:dtheta20_ev}
		\end{figure}

However, if the angle between both lattices is finite, super-lattice structures appear after a short time (see Fig.~\ref{fig:moire_density}(b)-(d)), which have a structure cell size that decreases for increasing values of $\theta_\Delta\in[0,\pi/6]$ and increases for larger values until the misalignment angle reaches the lattice symmetry point again at $\theta_\Delta=\pi/3$. These structures are transient and several revivals can be observed before the condensate settles back into the vortex lattice structure with an increase in the background wave-number spread, as expected based on kicking the non-rotating condensate. An example of this for a fixed angle is shown in Fig.~\ref{fig:dtheta20_ev}.	
		
To explain the interference patterns observed for misaligning the optical and the vortex lattice, we employ moir\'e interference theory \cite{SS:Kerman_jphyscon_2012}. Moir\'e patterns are known to appear when two periodic structures are overlaid while slightly misaligned to each other and can be calculated from the reciprocal lattice vectors. In all generality, any choice of equidistantly separated reciprocal lattice vectors be parameterised as
	\begin{equation}
		\mathbf{g}_{l} = g_0 \left[ \sin\left( \frac{2\pi l}{\alpha}+\theta \right),\, \cos\left( \frac{2\pi l}{\alpha} +\theta\right) \right],
	\end{equation}
where $\alpha$ describes the rotational symmetry of the lattice, $l$ labels the vector direction on the unit circle, $\theta$ is the angle with respect to a chosen coordinate system and $g_0$ is the reciprocal lattice constant. For our commensurate and triangular lattices we have $g_0=4\pi/(\sqrt{3}a)$, $\alpha=6$ and the vector directions are $l=\left[0\dots\alpha-1\right]$. As only the relative mis-alignement between the vortex and the phonon lattice matters, we choose $\theta=0$ for the vortex lattice and $\theta=\theta_\Delta$ for the optical potential alignment. 
	All possible wavelengths that can appear in an interference pattern between two such lattices in real space are then given by	
	\begin{equation}
		\lambda_{ll'} = \frac{\lambda_0}{|\mathbf{\mathbf{g}_{ll'}|}},
		\label{eq:InterferenceVectors}
	\end{equation}
	where 
	$\mathbf{g}_{ll'}=\mathbf{g}_{l}^\text{V}-\mathbf{g}_{l'}^\text{P}$, and 
	$\lambda_0 = 4\pi/\sqrt{3}$ for our commensurate triangular lattices.
One can see from Fig.~\ref{fig:dtheta20_ev} that a pattern matching the longest wavelength, $\lambda_M= \max[\lambda_{ll'}] \approx 2.9 a$, appears around $t=24$~ms and is clearly the most visible one. While patterns with shorter wavelength exist, they are harder to discern in our system and we therefore concentrate on the lowest wave-number in the following.

\begin{figure}[tb]
		\includegraphics[width=0.48\textwidth]{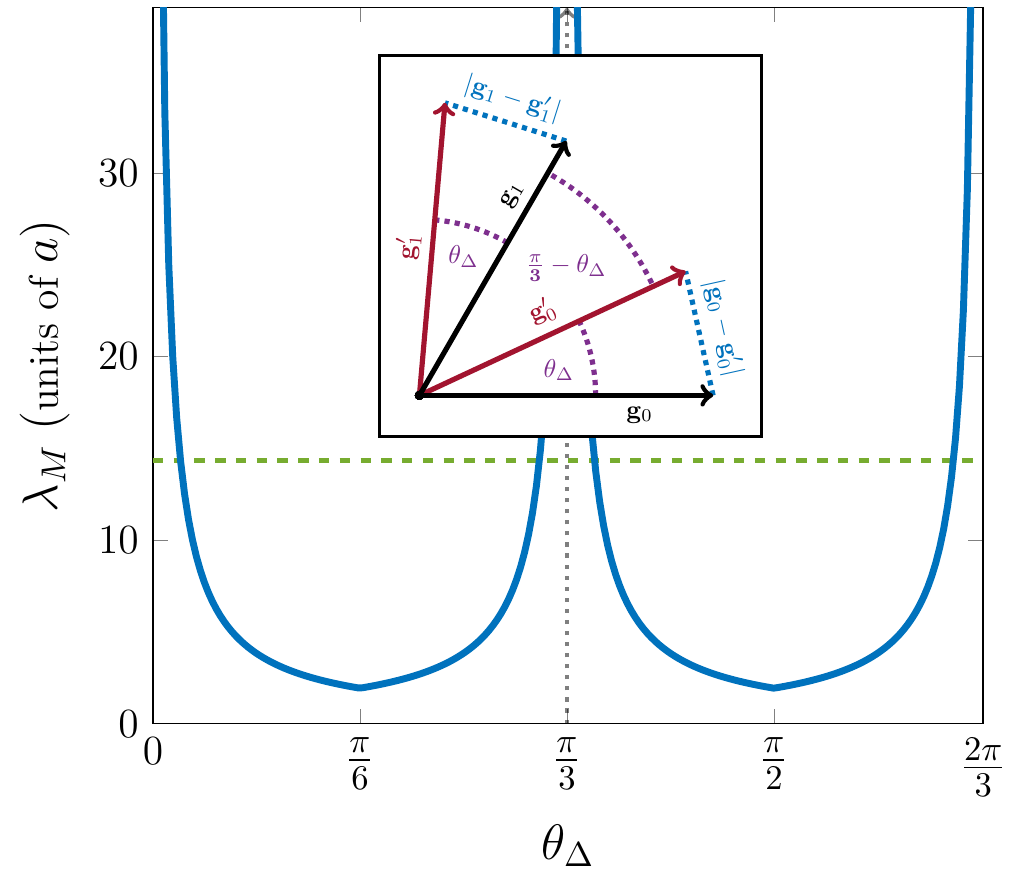}			
		\caption{Size of the resulting moir\'e super-structures as a function of the relative angle between the vortex and optical lattice. The dashed green line indicates the condensate radius. Inset: The different vectors in $\mathbf{k}$-space of the two lattices, with the optical lattice rotated by an angle $\theta_\Delta$. The $\mathbf{g}_{ll'} = |\mathbf{g}_l - \mathbf{g}_l'|$ vectors defining the dominant moir\'e wavelength are those for which the enclosed angle is smallest. }
		\label{fig:moire_lambda_1}
	\end{figure}

In $\mathbf{k}$-space the shortest $|\mathbf{g}_{ll^\prime}|$ corresponds to adjacent wave-vectors with the smallest $\theta_\Delta$ between them (see inset in Fig.~\ref{fig:moire_lambda_1}). Due to the symmetry of the lattices the most visible structures are therefore given by $\lambda_M=\lambda_{00}$ for $\theta_\Delta\in[0,\pi/6]$ and $\lambda_M=\lambda_{01}$ for $\theta_\Delta\in[\pi/6,\pi/3]$ (see inset of Fig.~\ref{fig:moire_lambda_1}).
While this symmetry assumption no longer holds strictly true after the system has been kicked, it is still fulfilled to a very good approximation during the initial dynamics. We can then
	obtain the wavelength of the dominating moir\'e structure as~\cite{jns-moire,nphys2272} 
		\begin{equation}
		\lambda_M = \frac{a}{2\sin(\eta/2)},
		\label{eqn:moire_size}
	\end{equation}
where $\eta=\min\{\theta_\Delta,\frac{\pi}{3} - \theta_\Delta \} $  (see Fig.~\ref{fig:moire_lambda_1}). 
	These super-structures 
	become observable when the wavelength becomes smaller than the radius of the condensate, which is $\lambda_M \approx 11a$
	and which corresponds to an angle $\theta_\Delta \approx \pi/36$.
	 One can see from Fig.~\ref{fig:moire_lambda_1} that once the relative angle is increased beyond this value  
	the structure sizes shrink to a minimum value at the point of complete misalignment, $\theta_\Delta=\pi/6$, giving $\lambda_M\approx 1.93\,a$, and increase again up to the point of symmetry. Beyond this point the whole behaviour starts over, due to the symmetry of the lattice. 
Note that in principle the above procedure can be carried out for square or other optical lattice geometries.

\begin{figure}[tb]
			\includegraphics[width=0.48\textwidth]{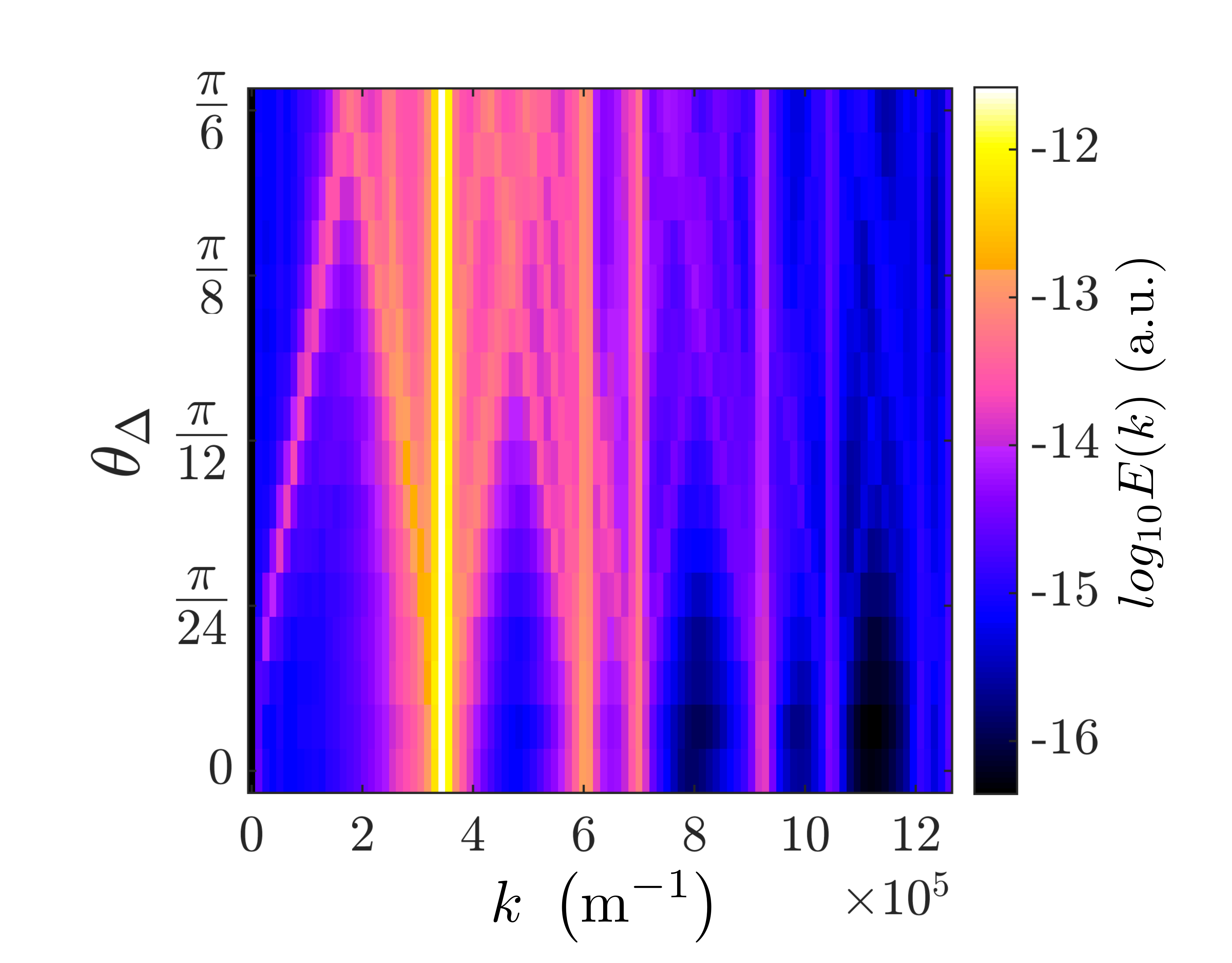}
			\caption{Compressible kinetic energy spectrum as a function of $\theta_\Delta$. All values are time-averaged over an interval $t=0$ s to $t=1$ s. The moir\'e peak corresponding to the lowest wave-number can be seen shifting to larger values for increasing angles and similar behavior is visible for the higher order components.}	
			\label{fig:dtheta_kspec}
		\end{figure}

The appearance of the moir\'e vector in $k$-space can be confirmed from the numerical simulations by looking at the compressible kinetic energy spectra which we display in Fig.~\ref{fig:dtheta_kspec}.

Apart from the dominant peaks corresponding to the underlying triangular geometry of the Abrikosov lattice, which are independent of $\theta_\Delta$ (straight lines in Fig.~\ref{fig:dtheta_kspec}), a number of additional peaks appear. Their position is a function of the misalignment angle and the lowest wave-number that appears increases its value with increasing $\Delta_\theta$. This is consistent with the moir\'e model and the appearance of density structures of differing size. 
Furthermore, a symmetric repeat of this structure about the $\theta_\Delta=\pi/6$ point is also visible, which corresponds to the $\pi/3 - \theta_\Delta$ lattice vector component. The minimum wavelength observed agrees with the theoretically determined minimum value of $\lambda_M\approx 1.93\,a$ and all other values over the range of observed angles. Note that for the higher harmonics at larger wave-numbers similar behaviour exists and is also covered by the moir\'e model.

\begin{figure}[tb]
			\includegraphics[width=0.48\textwidth]{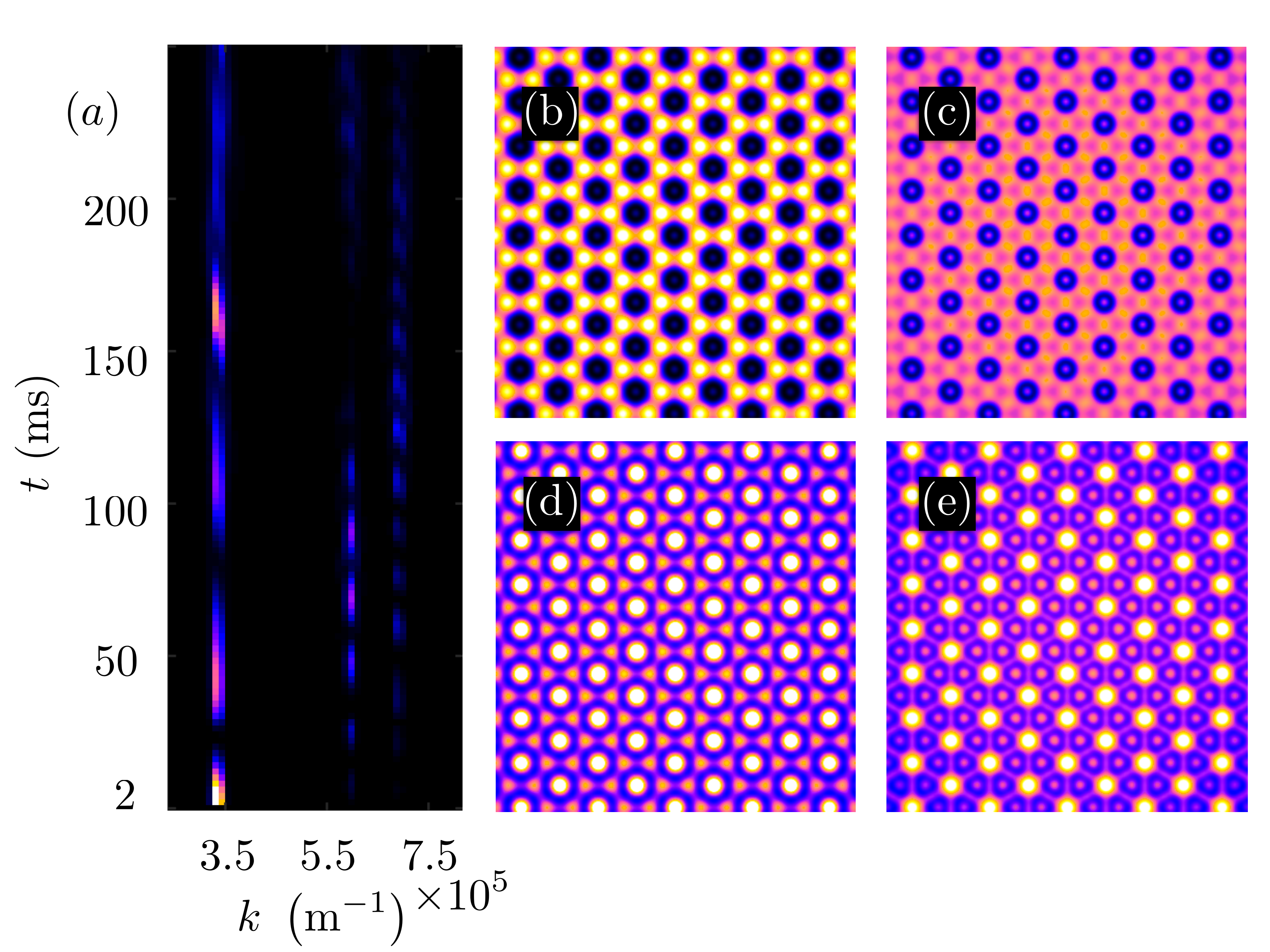}
			\caption{(a) For a kicking strength of $V_0 = 5.4\times10^{-2}\mu$ for a non-rotating condensate higher order modes become non-negligible contributions to the compressible kinetic energy spectrum. This leads to a variety of different density structures, with some close-ups shown for (b) 24 ms, (c) 36 ms, (d) 56 ms, and (e) 88 ms. Note that the larger structures in these plots are given by the optical lattice constant, which sets the scale.}
			\label{fig:kickp20k}
		\end{figure}

Let us briefly discuss what happens for stronger kicking, or when the two lattices are non-commensurate.
In the above the strength of the kicking pulse was chosen such that its perturbation only leads to a phase imprinting \cite{Vtx:Dobrek_pra_1999,BEC:Denschlag_science_2000}, with minimal change to the initial density. If one increases the kicking intensity the situation becomes quite different and one can see from Fig.~\ref{fig:kickp20k}(a) that higher order wave-numbers become more strongly excited. This, in turn, leads to modulations of the condensate density at shorter wavelength and an example is shown in Figs.~\ref{fig:kickp20k}(b)-(e). However, as for fully realistic experimental situations it is necessary to consider the heating of the condensates due to the disturbance once the kicking becomes stronger, we restrict this investigation to low intensity pulses.

A situation where the optical and the vortex lattice have different lattice constants can be imagined to appear naturally due to experimental uncertainties. Defining $a_o = a_v(1+\epsilon)$ the expression in eq.~\eqref{eq:InterferenceVectors} can be calculated to be 
\begin{equation}
	\lambda_M = \frac{a_v(1+\epsilon)}{\sqrt{2(1+\epsilon)(1-\cos\theta) + \epsilon^2}}
	\label{eqn:moire_size_eps}
\end{equation}
which reduces to eq.~\eqref{eqn:moire_size} for $\epsilon=0$. Evaluating this expression shows that the largest moir\'e wavelength changes slightly for small values of $\epsilon$, but it remains distinct enough from the underlying wave-vectors to stay visible in the evolution. This ensures that the system examined here is experimentally realistic.


	\section{Conclusions}
	\label{sec:Conclusions}
	
	We have examined the evolution of the density and kinetic energy spectra of a rapidly rotating Bose--Einstein condensate following a kick from a periodic optical potential. The kick was chosen to be short enough to effectively see a stationary condensate and weak enough to only lead to phase imprinting. The presence of an inhomogeneous phase  consequently leads to the appearance of phonon modes, which coherently interfere at certain times to form a lattice in the condensate density. This lattice, and the existing lattice from the Abrikosov arrangement of the vortices, were shown to interfere and give rise to so called moir\'e structures. As the system is dynamical, these structures are temporary and re-appear for a number of times before the spreading of the wave-numbers leads to the effect washing out.
	
	As the patterns that are created by the lattice interference are of much larger size than the underlying vortex lattice, the above mechanism of moir\'e interference can have an interesting application as a microscope. In our example the vortex lattice modifies the density on the scale of the healing length, which is not resolvable without time of flight data. However, the moir\'e-structures, which are a direct consequence of the lattice structure, are several times larger and might even be visible through direct imaging at the right time. This process could be applied to other density perturbations as well, for example higher order dark soliton trains. It would also be interesting to apply it to other vortex lattice configurations \cite{McEndoo:09} or vortices in the turbulent regime.

\section{Acknowledgments}

The authors acknowledge support for this work from the Okinawa Institute of Science and Technology Graduate University. We are grateful to JSPS for partial support from Grant-in-Aid for Scientific Research (Grant No. 26400422).
	
	\bibliographystyle{unsrt}
	
	\bibliographystyle{unsrt}	
\end{document}